\documentclass[12 pt]{article}

\usepackage{amsmath,graphics,amsfonts,amsthm,amssymb,color}
\usepackage{cite}
\usepackage{subfigure}
\usepackage{url}
\usepackage{epsfig}

\title{A Unified Term for Directed and Undirected Motility in Collective Cell Invasion}
\author{Jason M.\ Graham and Bruce P.\ Ayati}

\begin{document}
\maketitle

\begin{abstract}
  In this paper we develop mathematical models for collective cell motility. Initially we develop a model using a linear
diffusion-advection type equation and fit the parameters to data from cell motility assays. This approach is helpful in classifying
the results of cell motility assay experiments. In particular, this model can determine degrees of directed versus undirected collective cell motility.  Next we develop a model using a nonlinear diffusion term that is able capture in a unified way directed and undirected collective cell motility. Finally we apply the nonlinear diffusion approach to a problem in tumor cell invasion, noting that neither chemotaxis or haptotaxis are
present in the system under consideration in this article.
\end{abstract}

\section{Introduction}

  A basic characteristic of many organisms is their capability for active movement known as motility. Cell motility is fundamental in many important and interesting biological phenomena such as morphogenesis, wound healing, and tumor invasion and metastasis\cite{chicurel,friedl2,friedl}. The association of cell motility with diseases, and in particular cancer, makes it of interest not just for basic science but also in medical research.

  Much of the experimental research into cell motility focuses on molecular or mechanical details and is carried out in such a way that it is difficult to
  form a unified picture of cell motion \cite{chicurel}. To complicate matters further, both individual cell motion and collective cell motility are important in many phenomena. For example, the conventional view of tumor metastasis is that individual cells detach from a primary tumor and migrate to other areas \cite{friedl2}. However, now it is known that sheets or clusters of cells that maintain cell-cell contacts and migrate collectively also influence tumor invasion and metastasis \cite{friedl2,friedl}. Spurred by experimental work and the surrounding difficulties, researchers have begun to develop mathematical and computational models of cell motility (see \cite{anguige,chicurel,hillen} and references therein).  Such models can aid experimentalists in hypothesis testing, experiment design,
  and the creation of a unified picture of cell motility. A survey of mathematical and computational models of cell motility can be found in the article \cite{mogilner}. It would also be valuable if such models could be used to provide insight into systems of interest in biomedical research such as tumor invasion.

   Mathematical models of individual cell motion such as in \cite{hillen,mogilner} are able to incorporate much of the
   mechanical details of migration and model the migration of cells in tissue in which motion in some directions may be restricted. For example, in \cite{hillen} Hillen derives mathematical models for three dimensional cell motion through temporally varying tissue networks using kinetic transport equations. In this case, the motion in certain directions is limited by the orientation of tissue or extra cellular matrix.  This type of motion is often referred to as mesenchymal motion, and direction of motility is influenced by external factors not the intrinsic behavior of the cell population. Two cases are studied in \cite{hillen}, cell motion at the individual cell level and at the population level. Hillen begins
   with a transport equation model for individual cell mesenchymal motion and then uses scaling arguments to derive a model for motion of a population of cells.

   In this work
   we consider the development of models of collective cell motility that can be applied to the case of a group of cells that undergo collective
   migration where the degree of directed motility is determined by the intrinsic behavior of the cell population. More precisely, we consider the epithelial-mesenchymal transition (EMT) which is characterized by the loss of cell-cell adhesion
   and the resulting effects on motility \cite{friedl2,hugo,thierry}. The work presented in this paper is motivated in part by experiments of the Stipp lab \cite{stipp2,stipp1,stipp3,drake} and in part
   by the possibility of applying the results to simple but useful models of tumor invasion such as in \cite{stein,swanson1,swanson2}. We note that the cell motility scratch assays by the Stipp lab include neither chemotaxis since no chemoattractant is introduced into the scratch.
   Likewise, for the models in \cite{stein,swanson1,swanson2} of spherical \emph{in vitro} tumors no chemotaxis or haptotaxis since no chemoattractants or macromolecules exist outside the boundary of the tumor. In particular, the model of Stein et al.\ \cite{stein} is developed for glioblastoma tumor spheroid in a three-dimensional \emph{in vitro}
   experiment.

\section{Biological Background}

  Cell adhesion, is the result of the interactions of certain cell receptors with adhesion molecules. For example, proteins know as cadherins have been shown to influence cell-cell adhesion of cancer cells \cite{friedl2}. In particular epithelial cells express the protein E-cadherin which is responsible for the tight junctions between cells that is characteristic of the epithelium. Collective cell migration
  depends on cell-cell adhesion, while migration of individual cells is often associated with the loss of cell-cell adhesion. Such a loss of cell-cell adhesion
  may be an effect of an epithelial-mesenchymal transition (EMT). ``An EMT is a culmination of protein modification and transcriptional events in response to a defined
  set of extracellular stimuli leading to a long term, albeit
  sometimes reversible, cellular change. Core elements of EMT
  include reduction of cell-cell adherence via the transcriptional
  repression and delocalization of cadherins (adherens
  junctions), occludin and claudin (tight junctions), and
  desmoplakin (desmosomes).'' \cite{hugo}. It is thought that in (epithelial) cancers an EMT is essential for invasion and metastasis.

  The Stipp lab has performed experiments known as scratch assays in order to study the effects of an EMT on collective motility properties of
  cells \cite{stipp1}. In a scratch assay, cells are cultured on a surface and then a scratch is made to create a gap between some of the cells.
  Cells are then filmed migrating to ``fill in the gap''. Software (NIH Image J 1.63 software) can then be used to track some number of the cells in order to generate data
  from which cell velocities can be determined \cite{stipp1}. That is, the experiments provide data about the individual random walks of the cells. By applying standard statistical techniques to this data, one can compute diffusion and drift coefficients that offer a potential method for characterizing the results of such scratch assays. In particular, by computing the drift and diffusion coefficients we may be able to characterize the
  degree of cell-cell contact in a given assay. This is based on the observation that cell-cell contact influences motility properties (i.e. how quickly a population of cells migrates to fill in the gap).
  Using the mathematical models described below the motility properties can be characterized.

\begin{figure}
  \centering
  \includegraphics[width=2in,height=2in]{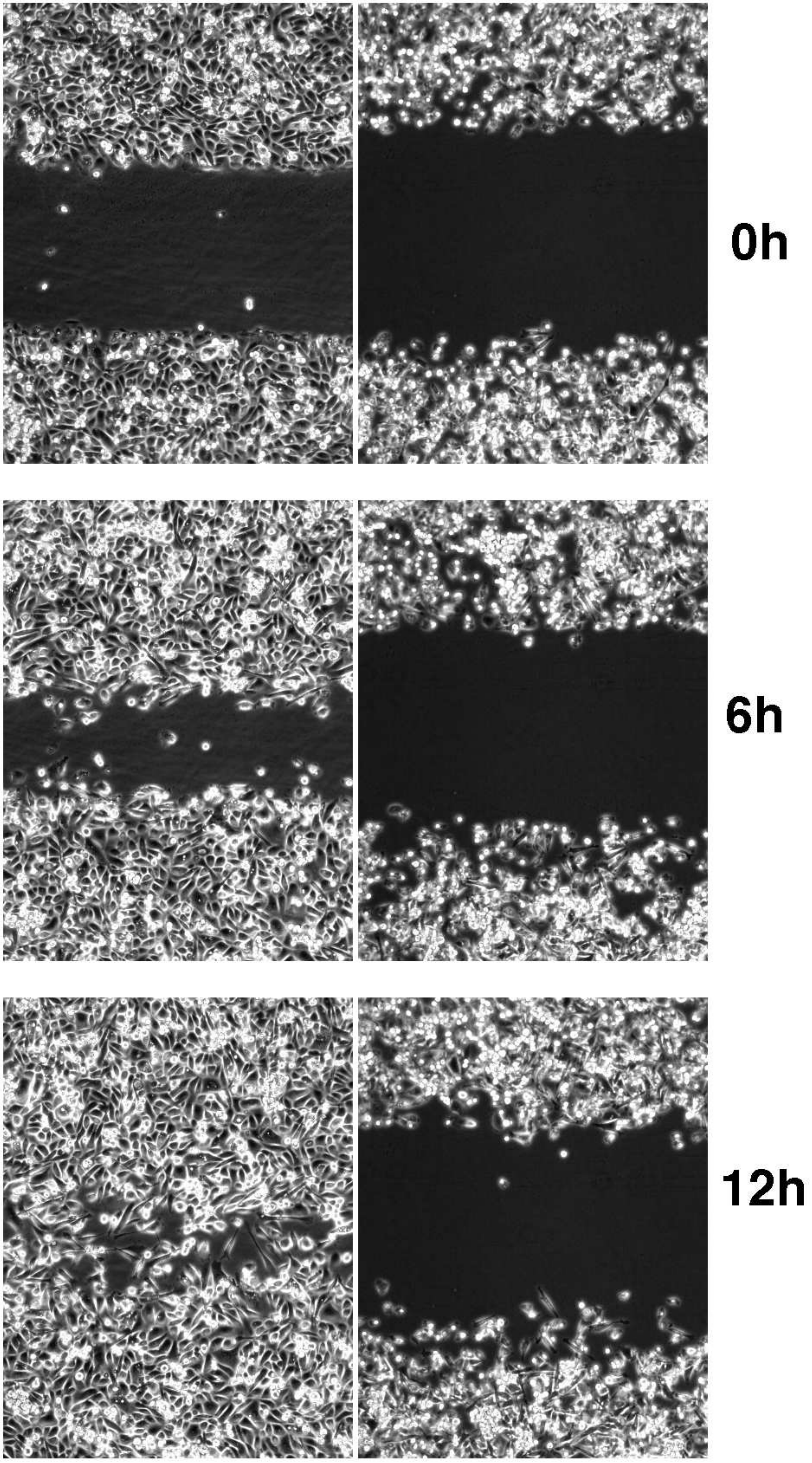}\\
  \caption{{\bf Scratch assay.} Figure shows stills at labeled times from a film of a scratch assay. Stills in the left hand column show PC-3 (epithelial) cells while the right column shows TEM4-18 (mesenchymal) cells. Note that cells on the right show reduced cell-cell adhesion. Figure courtesy of Christopher Stipp.}\label{gapPic}
\end{figure}

  Figure \ref{gapPic} shows frames from a film of a scratch assay performed by the Stipp lab. ``The cell types were PC-3 (epithelial) and TEM4-18 (mesenchymal).  TEM4-18 cells are a subline of PC-3 cells that was obtained experimentally by collecting a subpopulation of PC-3 cells with enhanced ability to traverse an endothelial cell layer. Later it was found that TEM4-18 cells had a gene expression profile consistent with a loss of epithelial cell characteristics, such as stronger cell-cell contacts, as compared with the parental PC-3 cell population as a whole. \cite{stipp1}'' The creation
  of the TEM4-18 subline from PC-3 cells is discussed in \cite{drake}.

\section{Mathematical Modeling}

  Using the data from the scratch assays, we have developed mathematical representations of collective motility. One such representation simply makes use of the diffusion and drift coefficients by input into a linear equation of the form
  \begin{align}
    \partial_{t}u & = \nabla \cdot (D\nabla u - {\bf v}u), \label{eq:driftDiff}
  \end{align}
  where $u(x,t)$ represents the cell density at a given position $x$ and time $t$, $D$ is the diffusion coefficient, and ${\bf v}$ is the drift velocity. An alternative approach developed in \cite{simpson2009} uses a discrete modeling method to build cell motility models based on individual cell random walk data.

  \begin{figure}
  \centering
  \includegraphics[width=2in,height=2in]{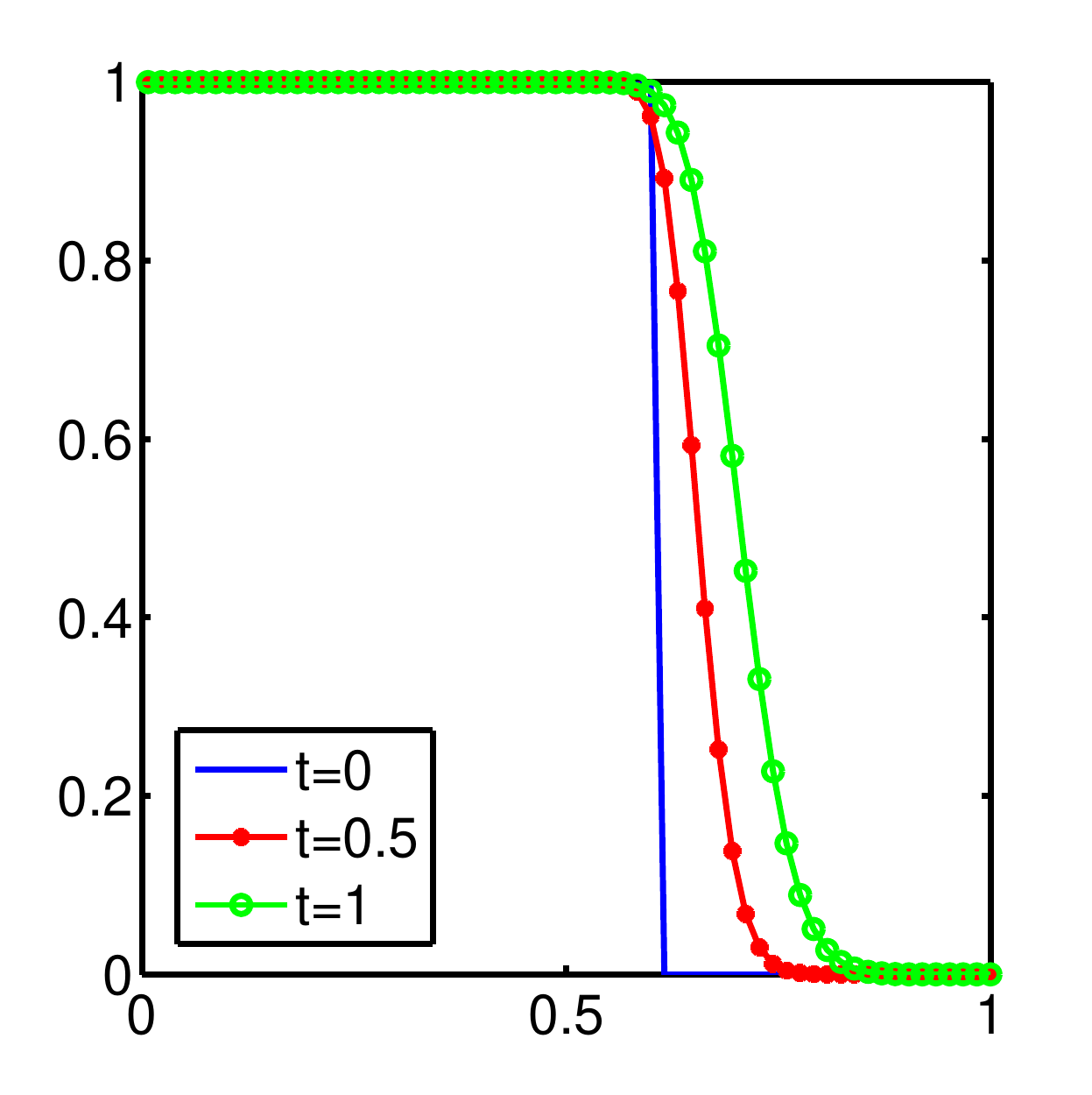}\\
  \caption{{\bf Results of the drift-diffusion model of cell motility.} This figure shows the front propagation of invading cells based on model (\ref{eq:driftDiff}) and data from scratch assays. The front propagation characterizes the motility properties of cells in a scratch assay. Since motility properties are linked with cell-cell contact, this allows one to characterize the degree of cell-cell contact by the drift and diffusion coefficient determined from the assay. }\label{driftDiffFig}
\end{figure}

  Figure \ref{driftDiffFig} shows the results typical of using (\ref{eq:driftDiff}) to track the front propagation in a scratch assay. Here
  we have used a scaling to have unit maximum cell population density. The spatial domain for the computation is the interval $[-15,1]$. On the
  left boundary we take $u(-15,t)=1$ for all $t$, a so-called far-field condition. This is justified, since in a typical scratch assay the scratch is small compared with
  the area containing cells. We employ a no-flux condition on the right boundary. We take as the initial condition
  \begin{align}
    u(x,0) & = \left\{\begin{array}{ll} 1, & \mbox{if $x \leq \frac{3}{5}$,} \\ 0, & \mbox{otherwise.}  \end{array} \right.
  \end{align}
  This represents the assay at the time of the scratch. Here we only consider cells to the left of a scratch as in figure \ref{gapPic}. This is sufficient due to the left-right symmetry about the center of the scratch. The scratch creates a front between a region with cells and a region without cells, which moves as the cells
  migrate to fill in the gap. Figure \ref{driftDiffFig} shows a right-moving cell population that corresponds to the motion of the front in the  scratch assays. We obtain from the computations the front speed and compare it with experimental results. In this case parameter values are chosen to correspond to a scratch assay with epithelial (PC-3) cells. This provides a method for characterizing the motility properties, and possibly the degree of cell-cell adhesion, in such assays.

  In \cite{aronson}, Aronson describes the role of diffusion equations in modeling dispersing populations. In particular, \cite{aronson}
  covers how the behavior of a dispersing population is related to the form of the diffusion operator appearing in the equations
  modeling the spatial dynamics of a population. It is pointed out that it can be useful to consider nonlinear diffusion terms. In \cite{anguige}
  the authors present a discrete model of cell motility that incorporates cell-cell adhesion. This discrete model has as its continuum limit
  a nonlinear diffusion equation of the form
   \begin{align}
      \partial_{t}u & = \nabla \cdot (D(u)\nabla u). \label{eq:nonlinDiff}
   \end{align}
   Due to the occurrence of cell-cell adhesion, in \cite{anguige}, the diffusivity is described by a function of the form
   \begin{align}
      D(u) & = a\left(u - b\right)^{2} + c, \label{eq:anguigeDiff}
   \end{align}
   where $a,b,c$ are constants related to the adhesion properties of cells. Under the influence of \cite{anguige,aronson}
   we seek an alternative to (\ref{eq:driftDiff}) by considering nonlinear diffusion equations of the form (\ref{eq:nonlinDiff}).

\begin{figure}
  \centering
  \includegraphics[width=2in,height=2in]{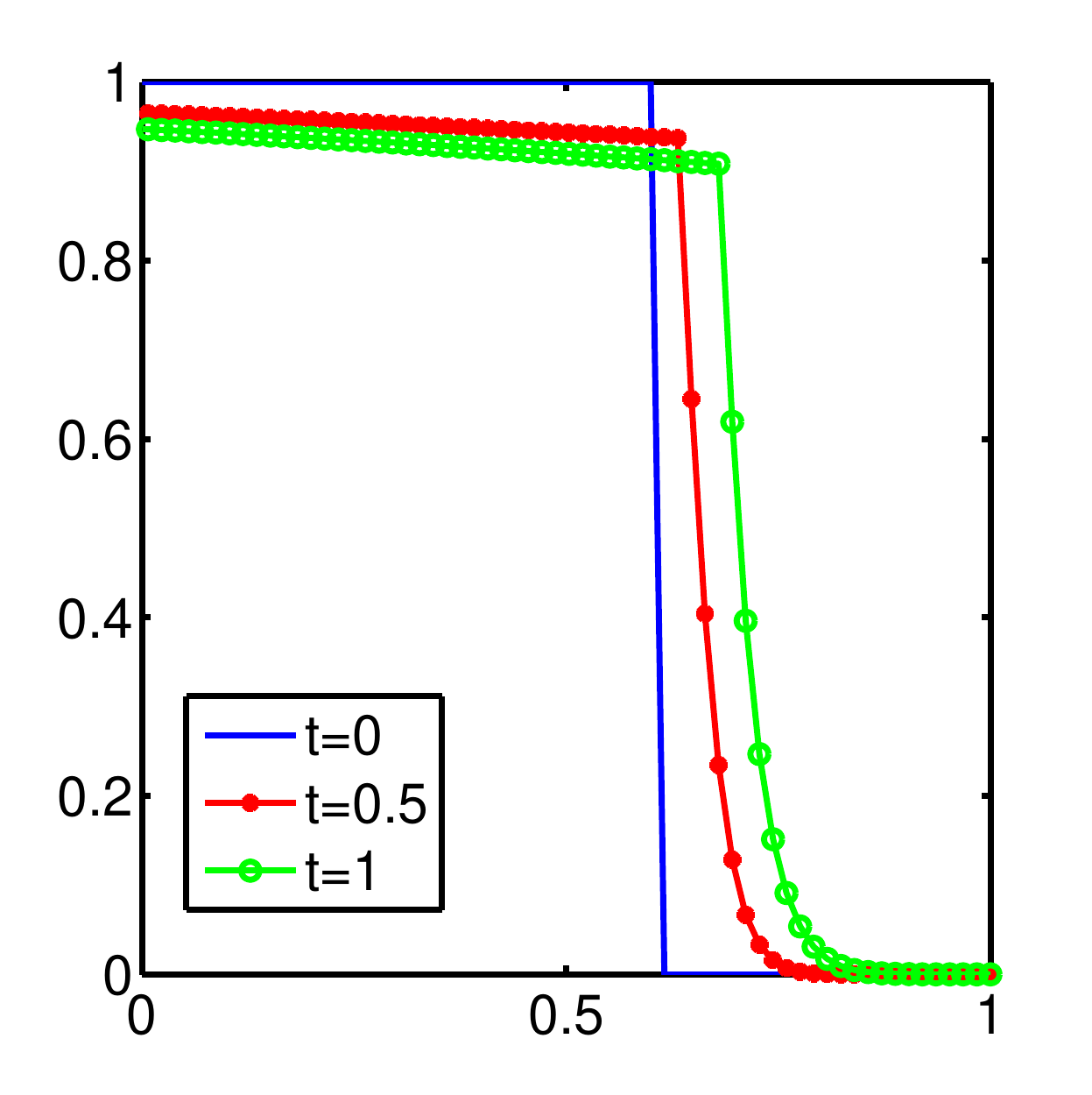}\\
  \caption{{\bf Results of nonlinear model of cell motility.} This figure shows the front propagation of invading cells based on model (\ref{eq:nonlinDiff}).}\label{nonlinDiffFig}
\end{figure}

   Figure \ref{nonlinDiffFig} shows the results of using a nonlinear diffusion model of the form (\ref{eq:nonlinDiff}) to represent
   cell motility as observed in the scratch assays discussed above. In order to get a qualitative match with the results, such as in figure \ref{driftDiffFig}, using (\ref{eq:driftDiff}) we have taken the nonlinear diffusivity to have the form
   \begin{align}
      D(u) & = a\cdot \text{max}(0,u-b)u + c, \label{eq:nonlinDiffCoeff}
   \end{align}
   which is similar to (\ref{eq:anguigeDiff}). The boundary and initial conditions are the same as above. Figure \ref{nonlinDiffFig}
   again simulates a right-moving cell population (epithelial (PC-3) cells) that corresponds to the motion of the front in the scratch assays.
   The advantage of using a nonlinear diffusion model is that the effects of the degree of cell-cell adhesion on motility, of interest in the
   study of the EMT, can be explicitly represented. Note that by making the constant $a$ smaller or the constant $b$ larger in (\ref{eq:nonlinDiffCoeff}) the diffusion becomes closer to linear, that is we get closer to purely random motility.

\section{Application to Tumor Invasion}

 In \cite{swanson1,swanson2} Swanson et al.\ apply simple but useful mathematical models of tumor invasion. In particular the authors use a
 reaction-diffusion equation of the form
 \begin{align}
    \partial_{t}u & = \nabla \cdot (D\nabla u) + gu\left(1-\frac{u}{u_{\text{max}}}\right) \label{eq:swan}
  \end{align}
 to model the invasion of a spherically symmetric gliobastoma tumor. The use of a reaction-diffusion equation is based on the assumption that the net dispersal of tumor
 cells is captured by a random walk which translates mathematically into diffusion. In the case of \cite{swanson1,swanson2} diffusion is assumed to be linear. The term $gu(1-\frac{u}{u_{\text{max}}})$ describes the proliferation of cells. Based on invasion assays of tumor cells, Stein et.\ al.\ \cite{stein} suggest that there is a need to consider not only
 random motility of invasive cells but also directed motility. In particular the authors of \cite{stein} conclude that while diffusion alone
 is sufficient for describing cell motility of cells making up the central core of the tumor, cells at the invasive rim of the tumor
 have different motility properties that include both random and directed motility. By modifying (\ref{eq:swan}) the authors of \cite{stein}
 arrive at the following model for the invasion of tumor cells
 \begin{align}
    \partial_{t}u & = \nabla \cdot (D\nabla u - {\bf v}u) + s\delta(x - x_{0} - {\bf v} t ) + gu\left(1-\frac{u}{u_{\text{max}}}\right). \label{eq:stein}
 \end{align}
 Here $\delta(\cdot)$ is the Dirac delta function.
 In the above model the $s\delta(x - x_{0} - {\bf v} t )$ term represents the shedding of invasive cells from the invasive rim.

\begin{figure}
   \centering
   \subfigure[t=0]{\includegraphics[height=1.5in]{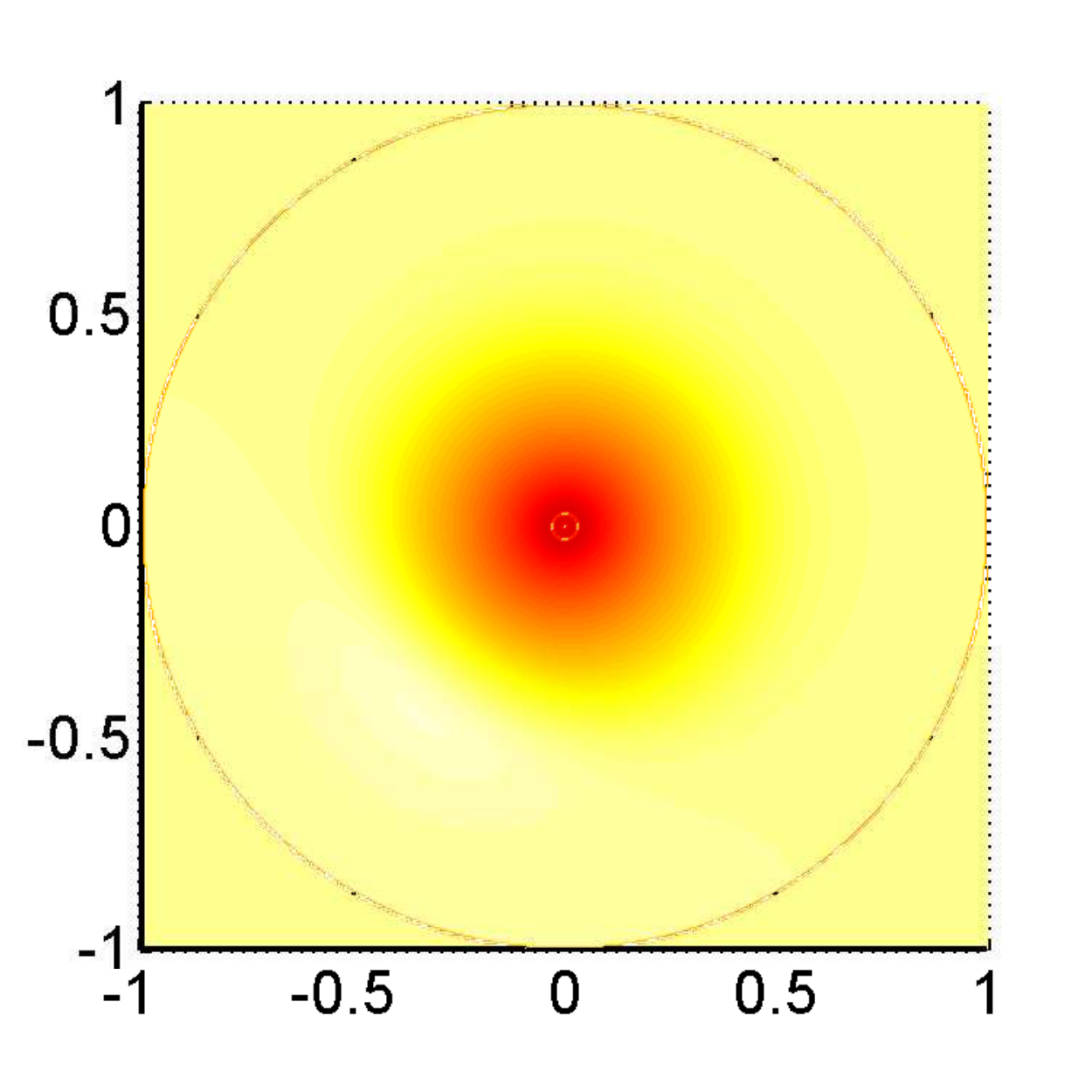}\label{stein-a}}
   \subfigure[t=1.25 days]{\includegraphics[height=1.5in]{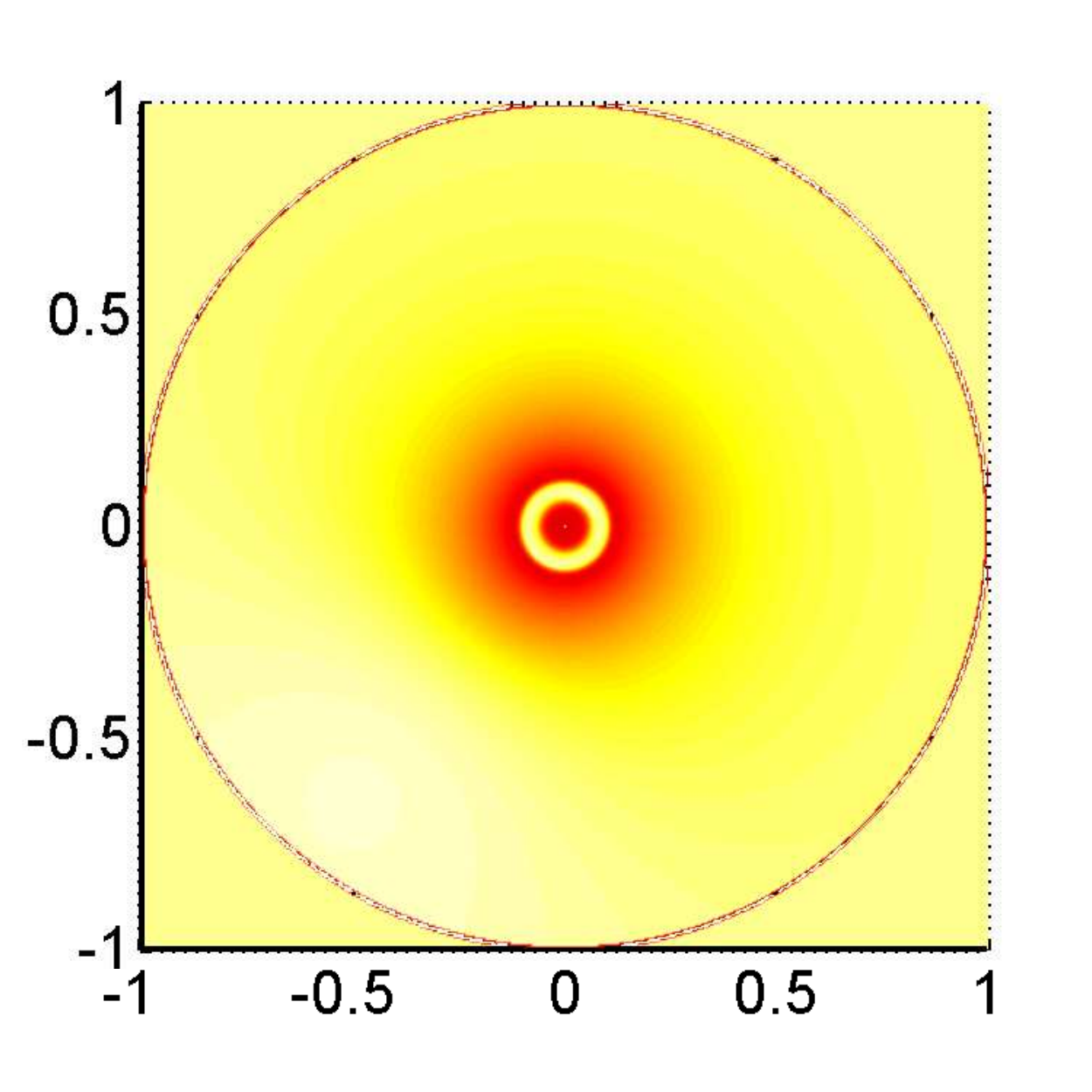}\label{stein-b}}
   \subfigure[t=2.5 days]{\includegraphics[height=1.5in]{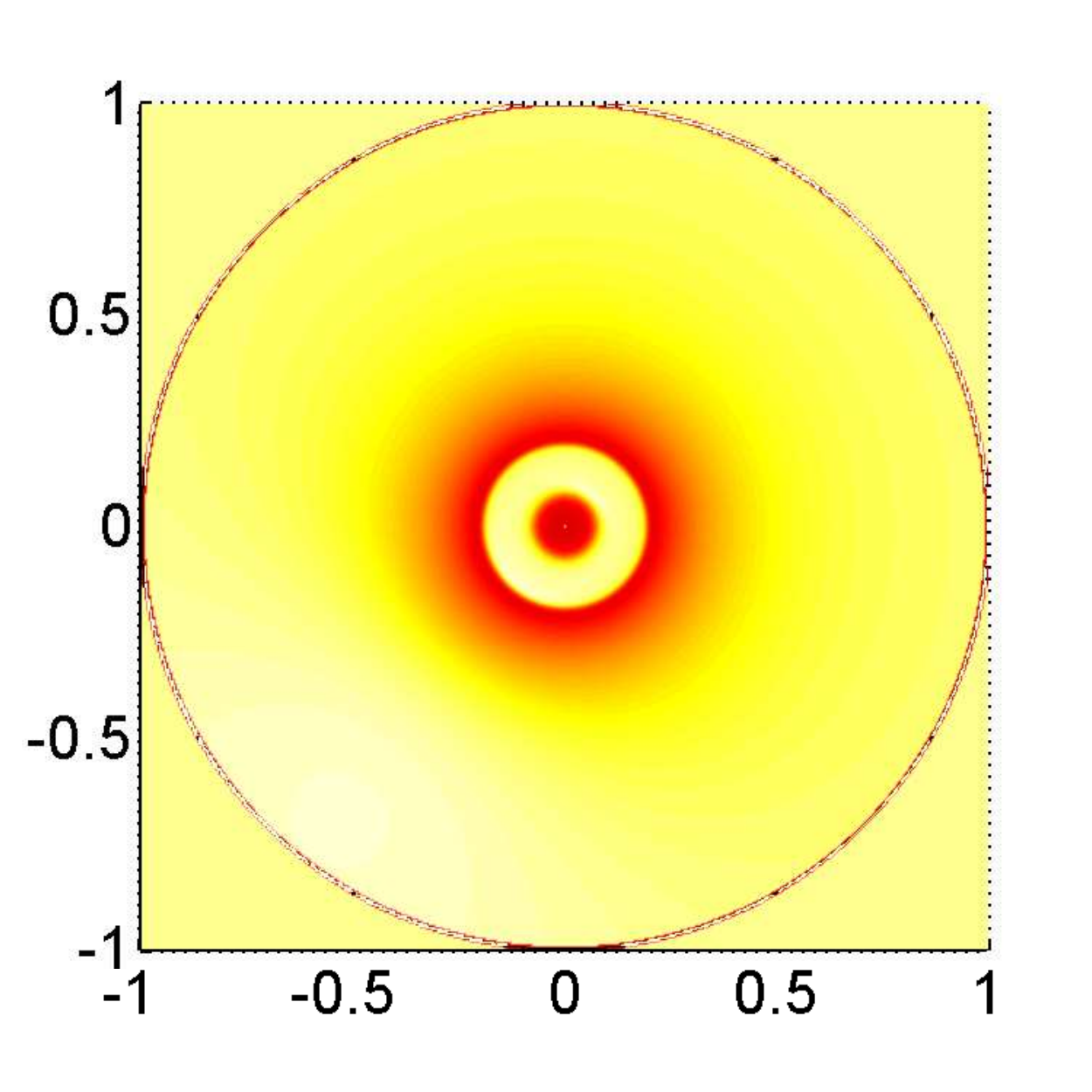}}\\
   \subfigure[t=3.75 days]{\includegraphics[height=1.5in]{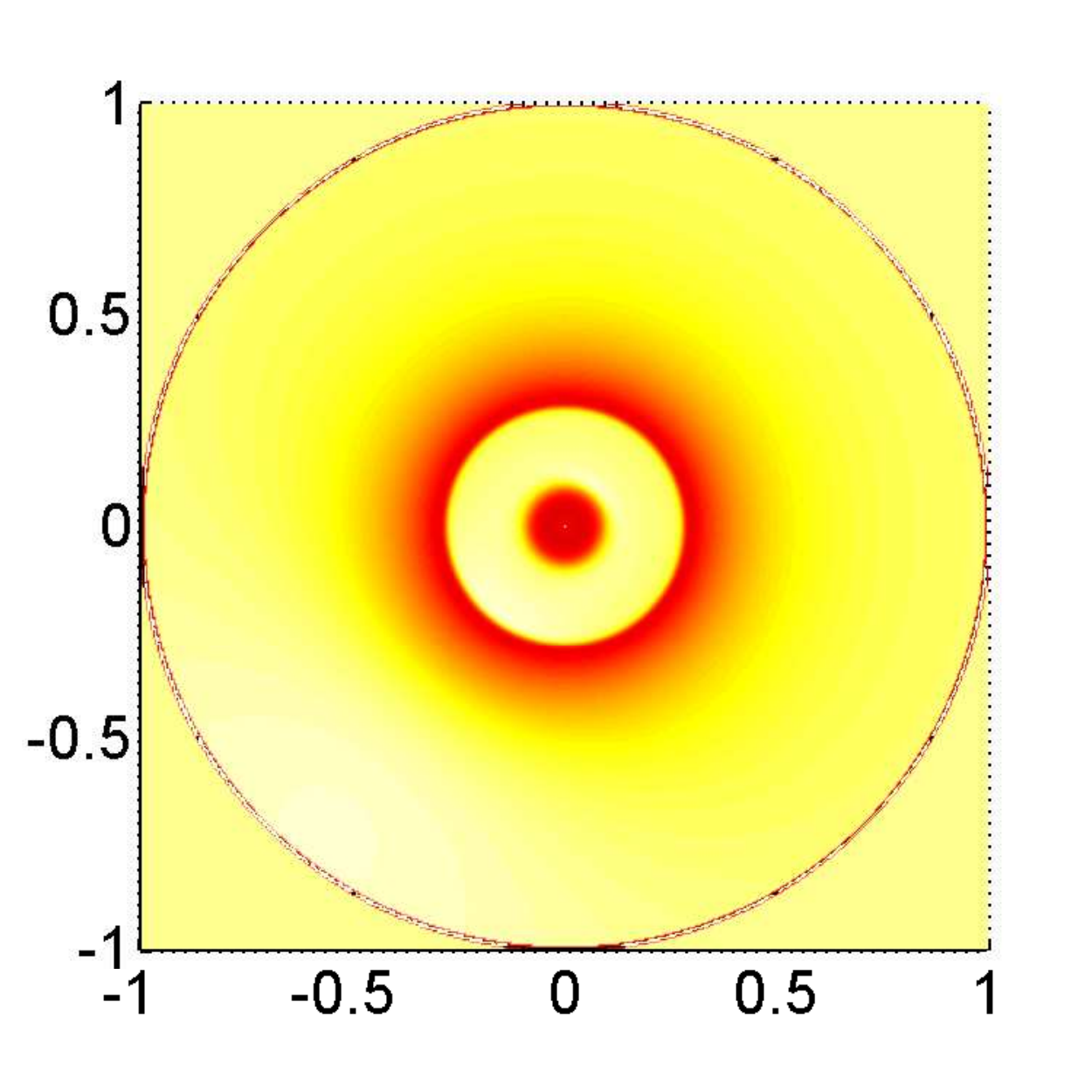}}
   \subfigure[t=5 days]{\includegraphics[height=1.5in]{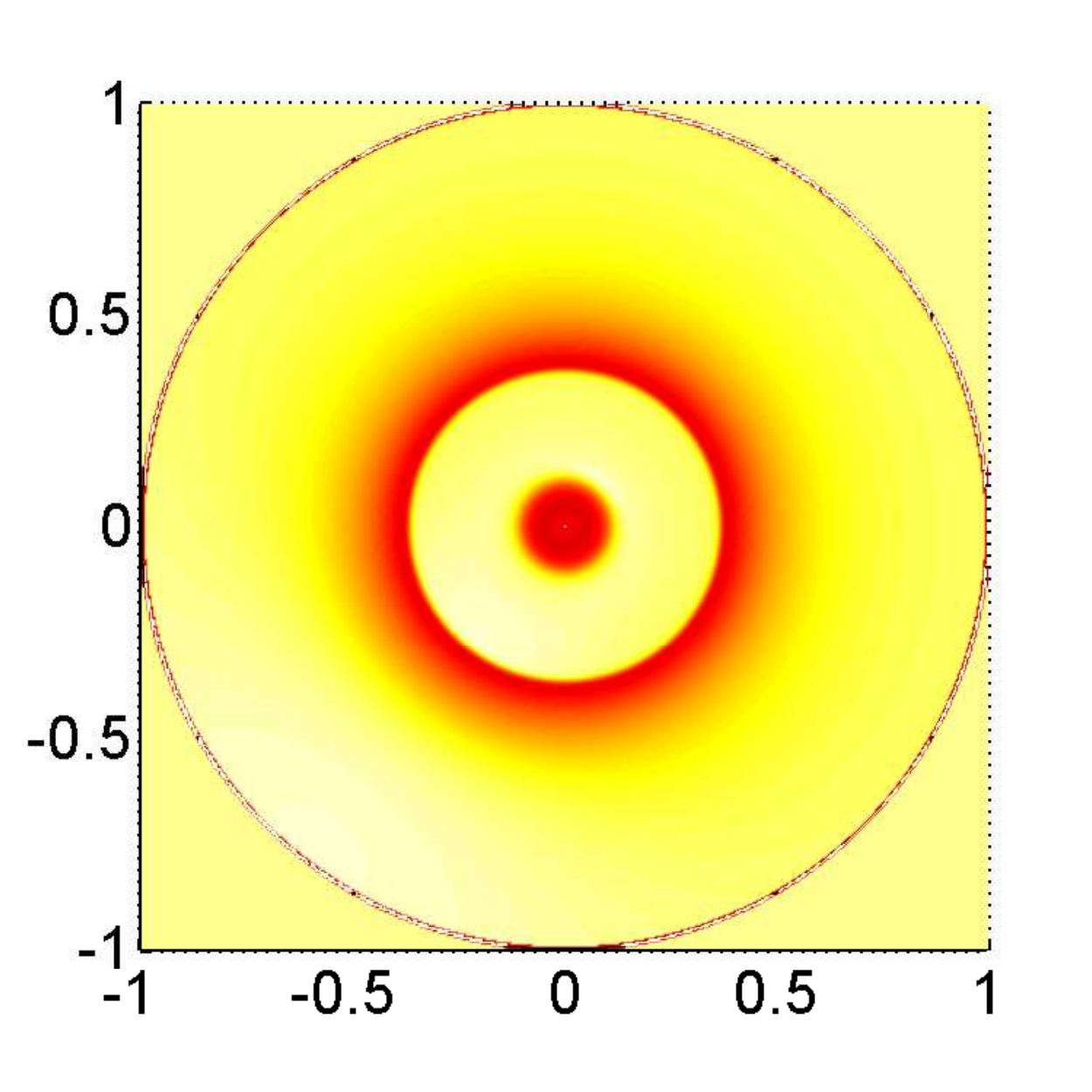}\label{stein-e}}
   \subfigure[t=6.25 days]{\includegraphics[height=1.5in]{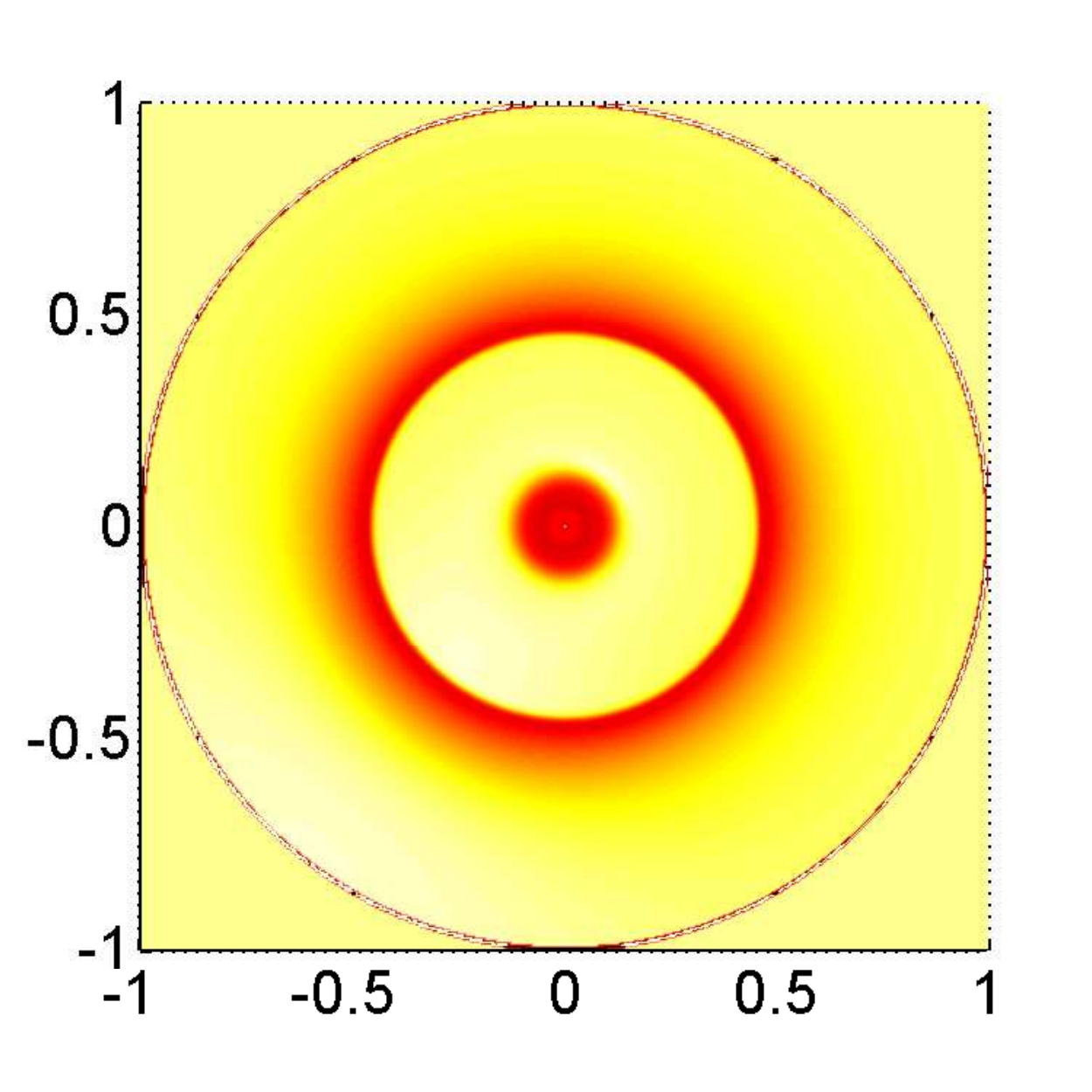}\label{stein-f}} \\
   \subfigure[t=7.5 days]{\includegraphics[height=1.5in]{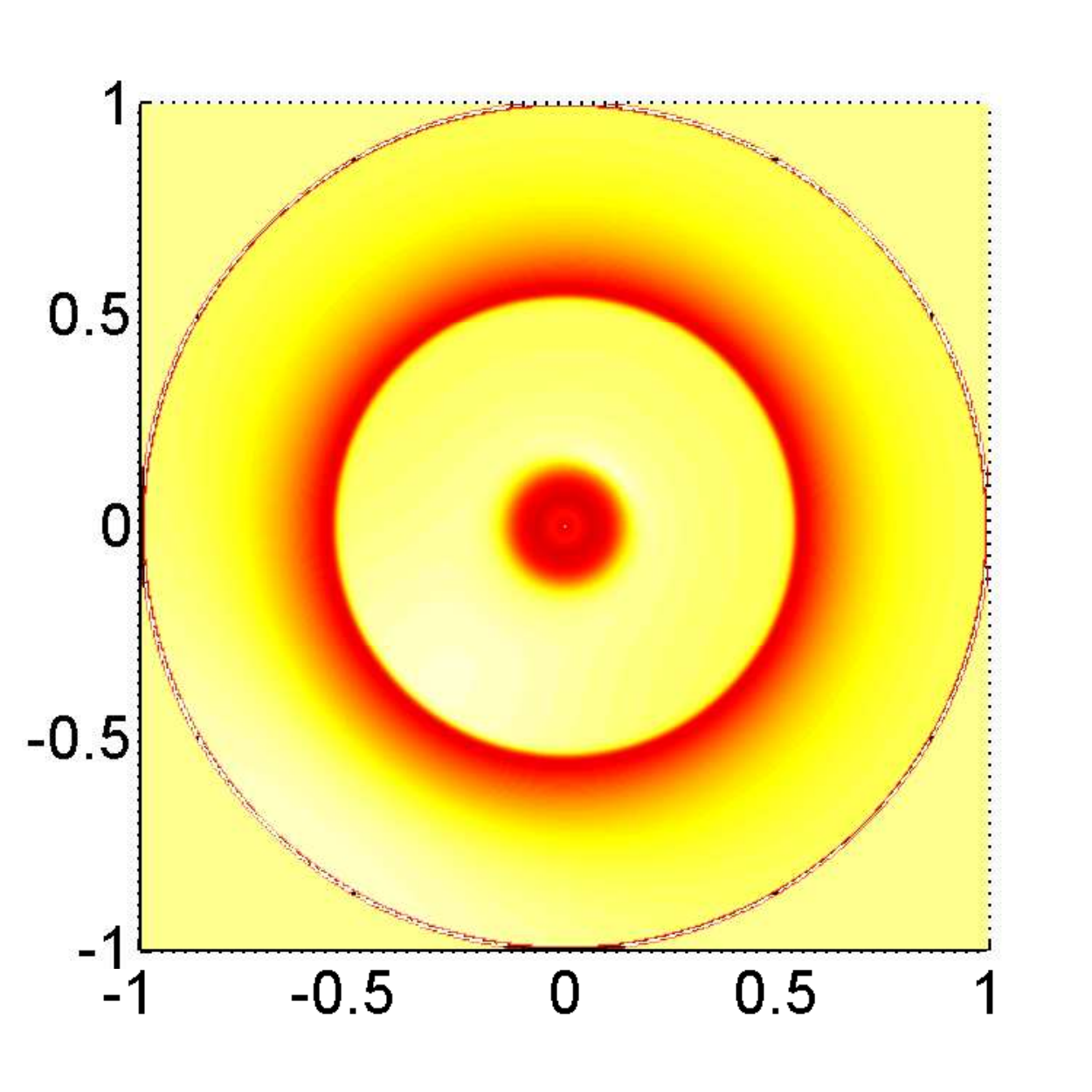}\label{stein-g}}
   \subfigure[t=8.75 days]{\includegraphics[height=1.5in]{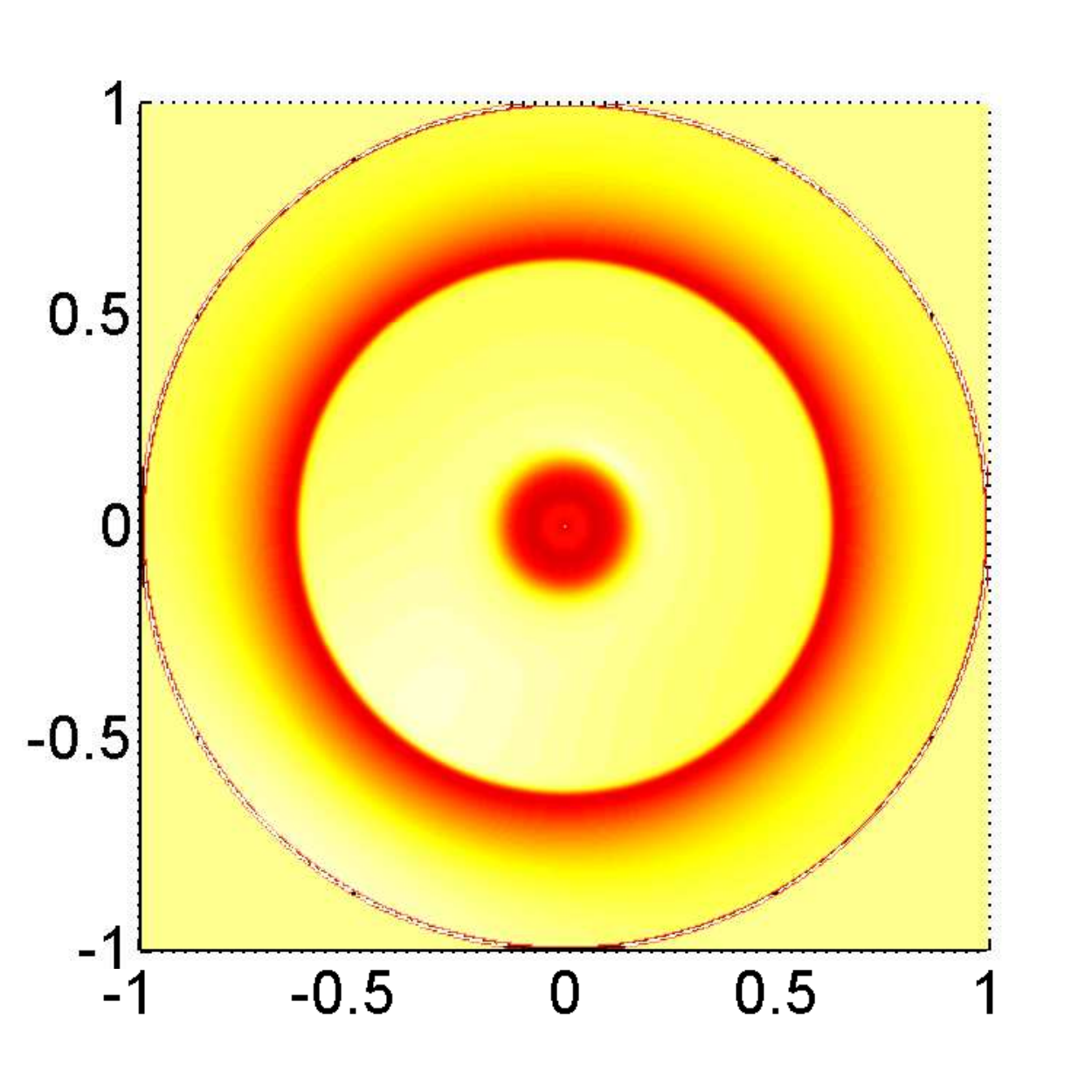}}
   \subfigure[t=10 days]{\includegraphics[height=1.5in]{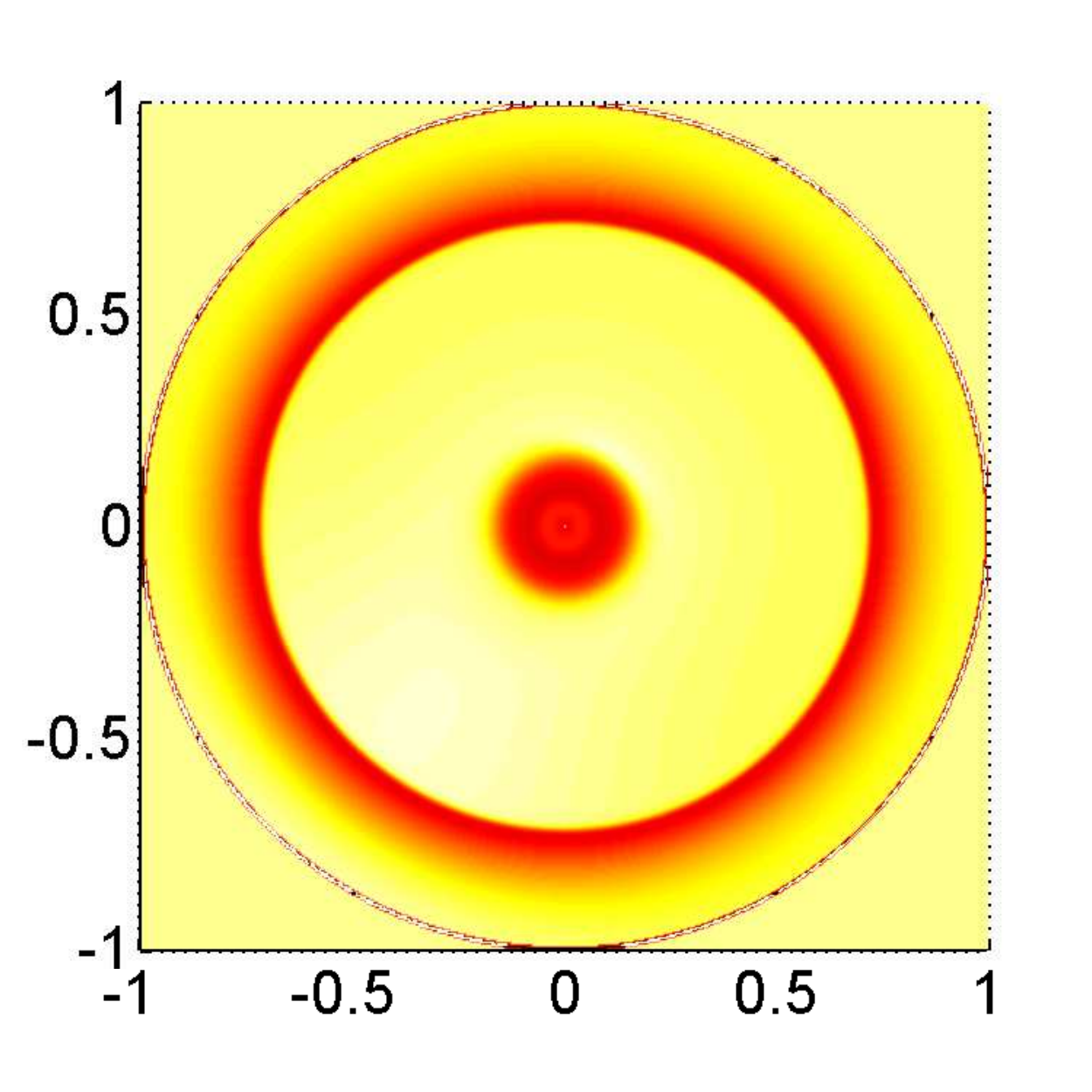}\label{stein-i}}
   \caption{{\bf{Simulation of tumor invasion with Stein et al.\ model.}} This figure shows simulations (using (\ref{eq:stein})) of invasion of a spherically symmetric tumor. The simulations are meant to model experimental results described in \cite{stein}. The parameter values from \cite{stein} are used.}
   \label{stein}
 \end{figure}

 Figure \ref{stein} shows simulations using (\ref{eq:stein}). Here we have assumed a spherically symmetric tumor as done in \cite{stein,swanson1,swanson2}. The interesting feature of these results is the behavior of the invasive rim. We note
 that in \cite{stein} the cells forming the tumor core are treated separately, as a result the behavior of the cells
 making up the core are not shown in figure \ref{stein}. In \cite{stein}, the parameters $D,{\bf v},s,g$ in (\ref{eq:stein}) are optimized
 to fit experimental results obtained by the authors.

\begin{figure}
   \centering
   \subfigure[t=0]{\includegraphics[height=1.5in]{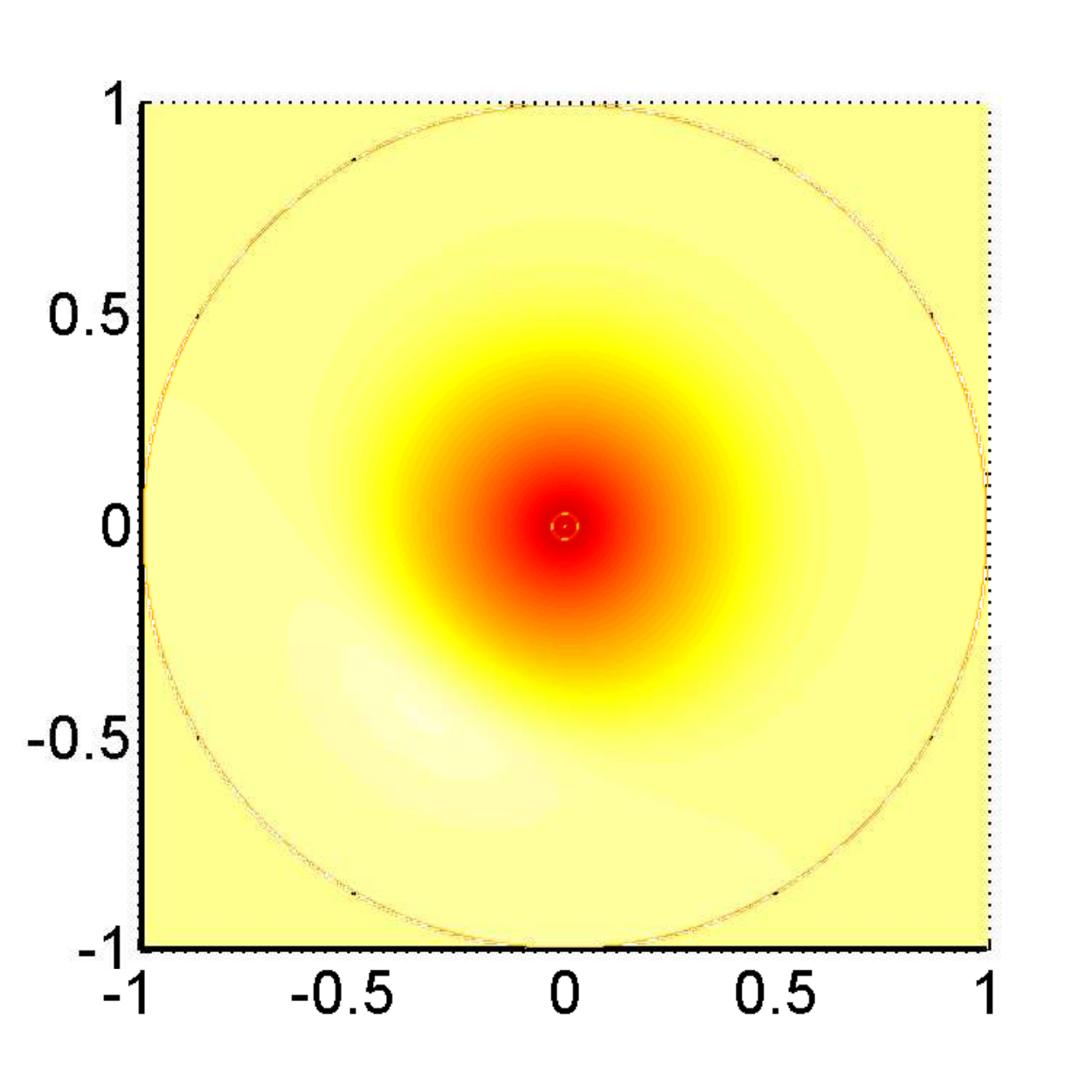}\label{if-a}}
   \subfigure[t=1.25 days]{\includegraphics[height=1.5in]{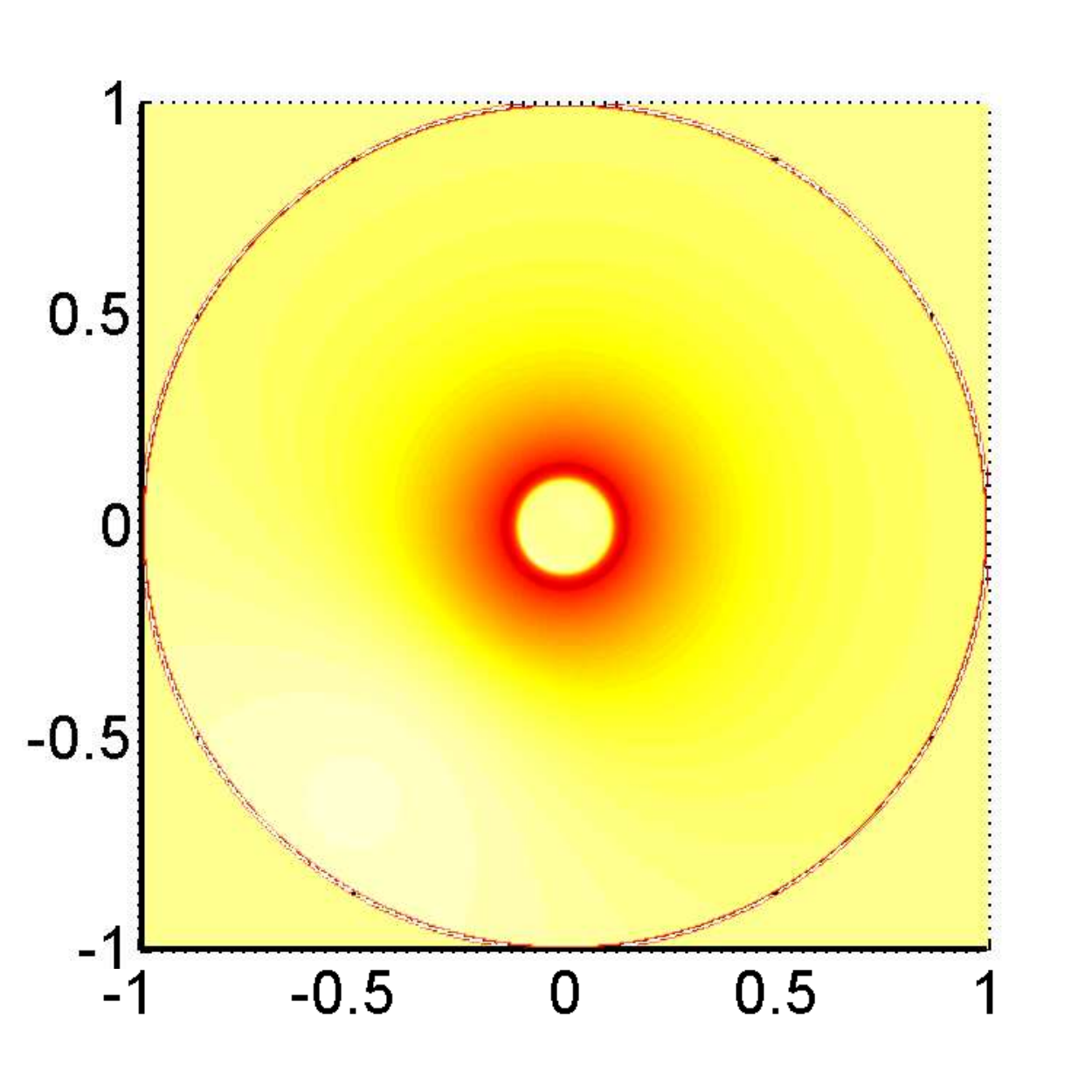}\label{if-b}}
   \subfigure[t=2.5 days]{\includegraphics[height=1.5in]{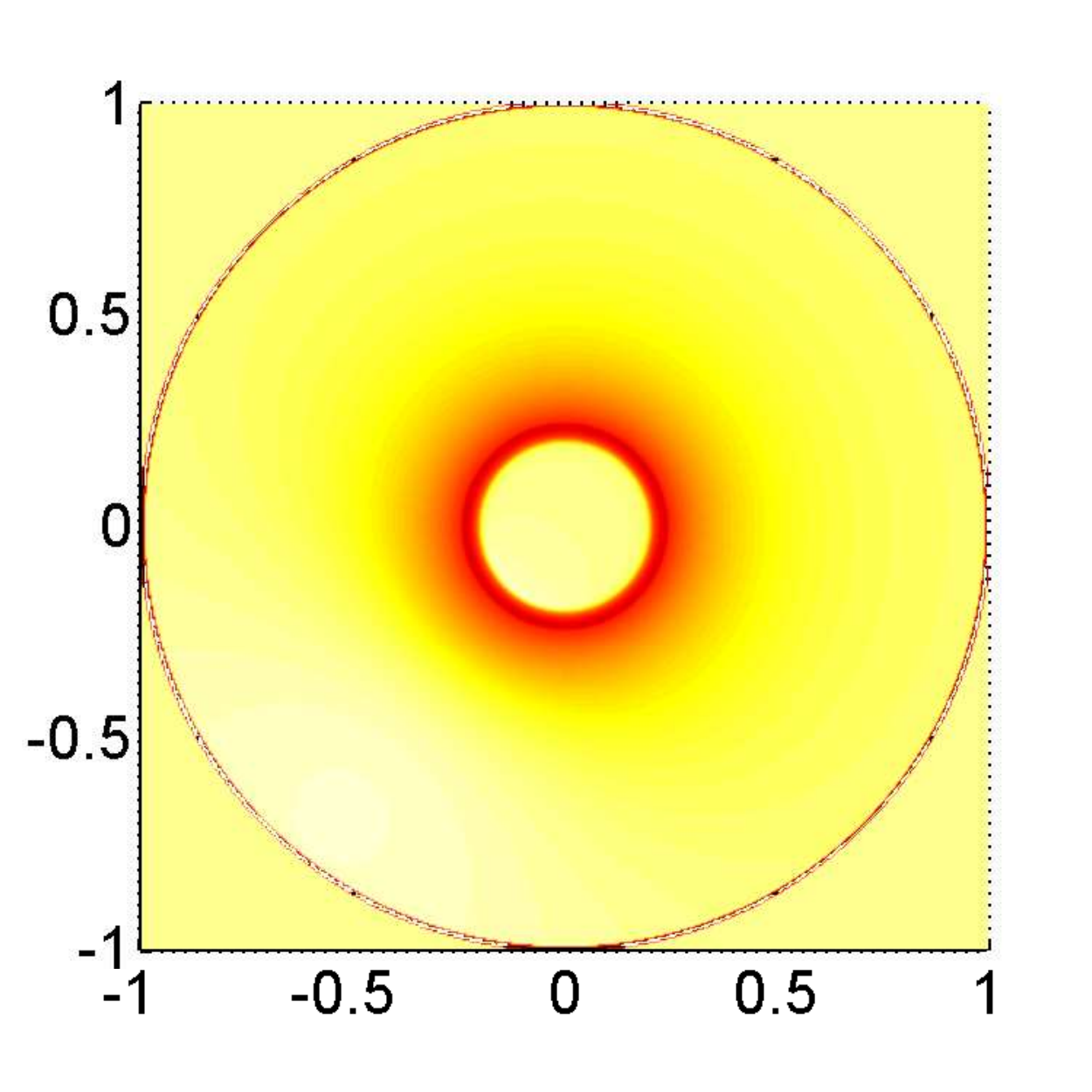}}\\
   \subfigure[t=3.75 days]{\includegraphics[height=1.5in]{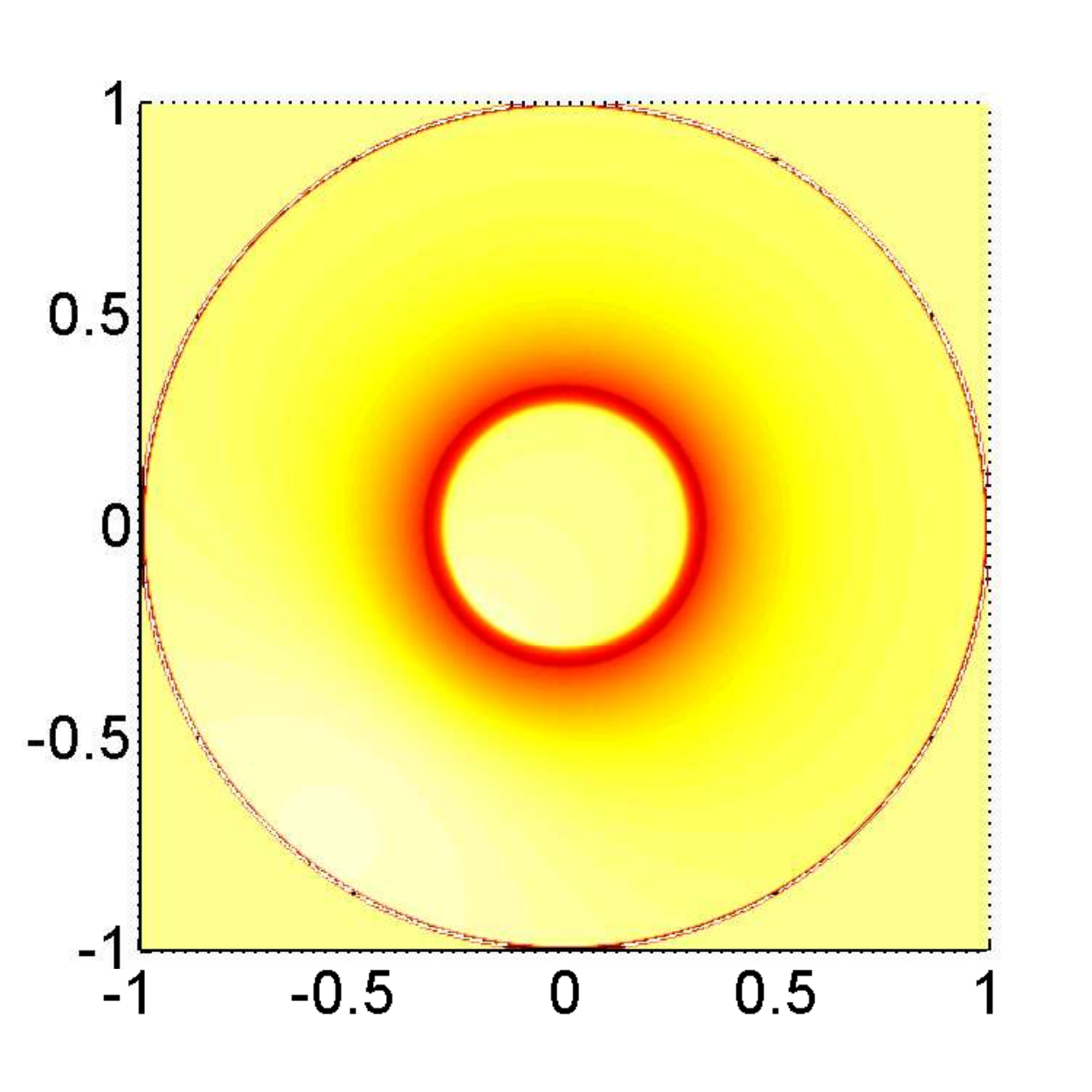}}
   \subfigure[t=5 days]{\includegraphics[height=1.5in]{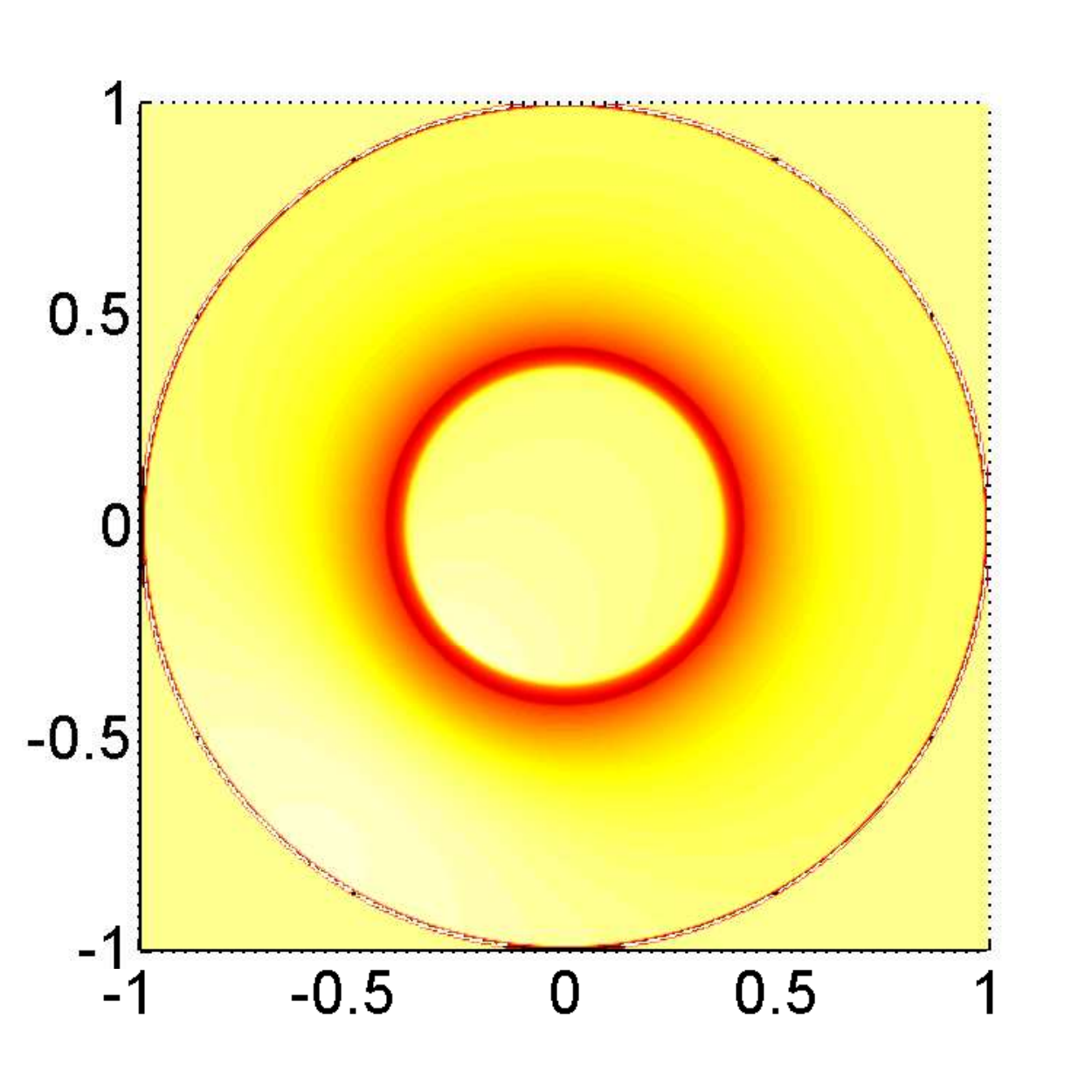}\label{if-e}}
   \subfigure[t=6.25 days]{\includegraphics[height=1.5in]{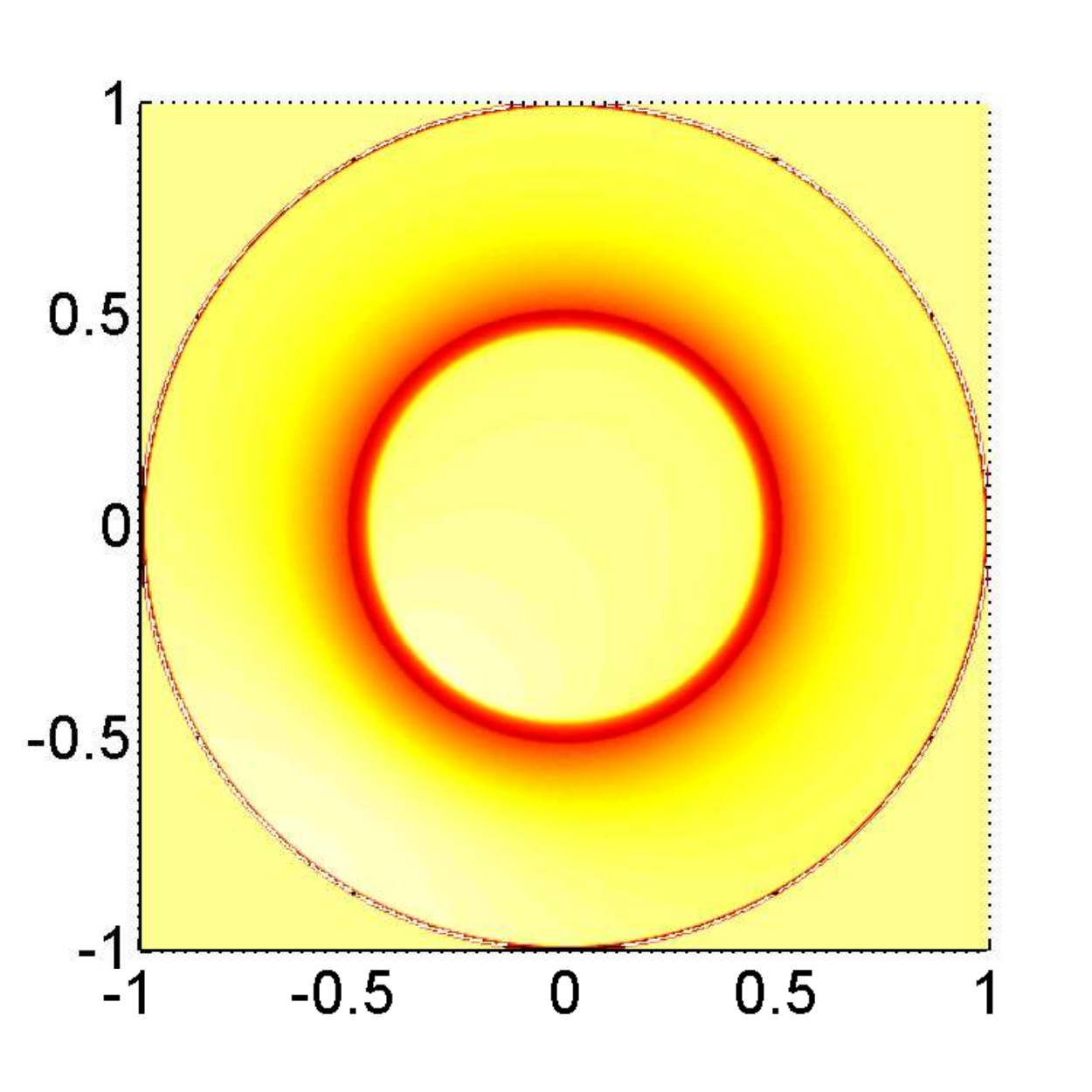}\label{if-f}} \\
   \subfigure[t=7.5 days]{\includegraphics[height=1.5in]{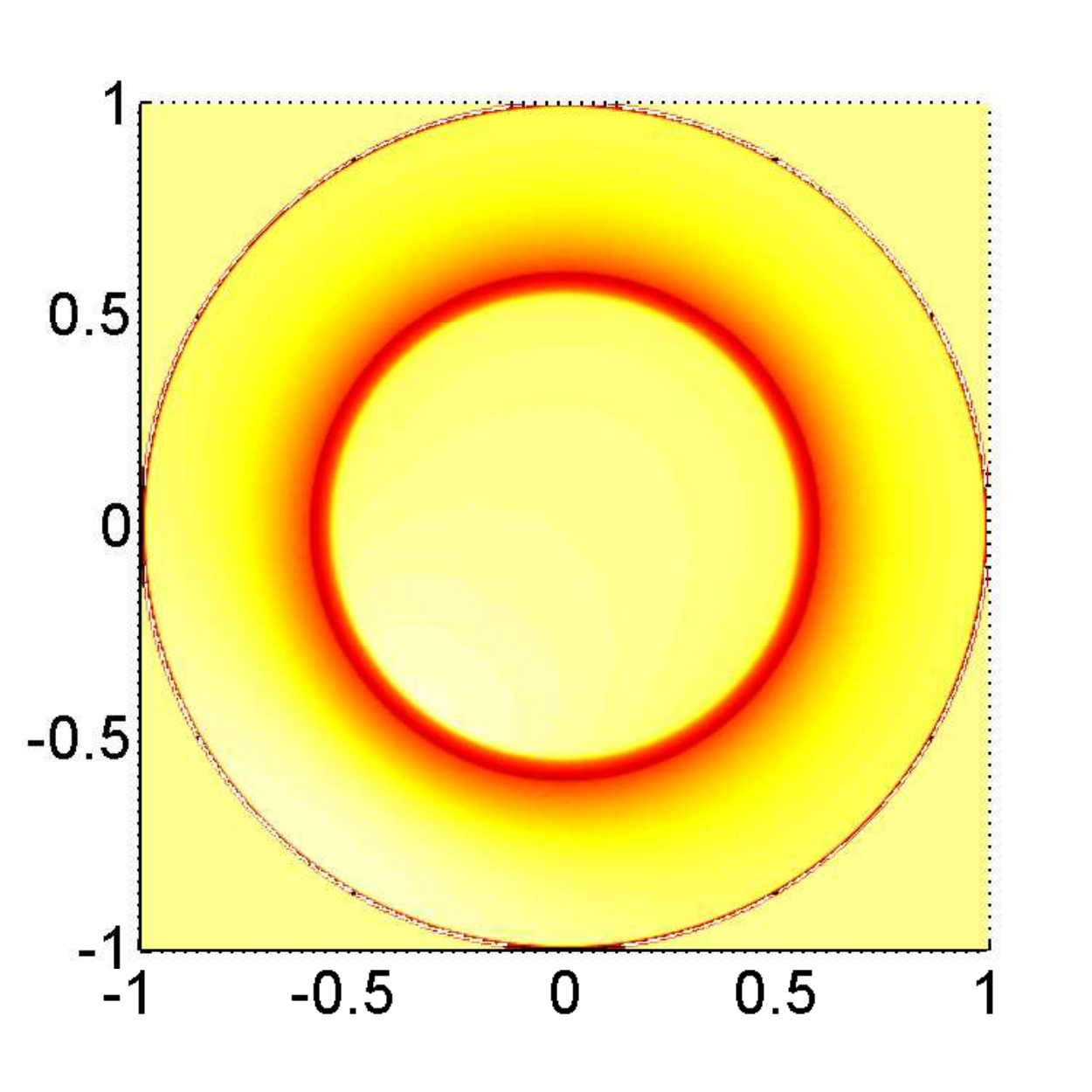}\label{if-g}}
   \subfigure[t=8.75 days]{\includegraphics[height=1.5in]{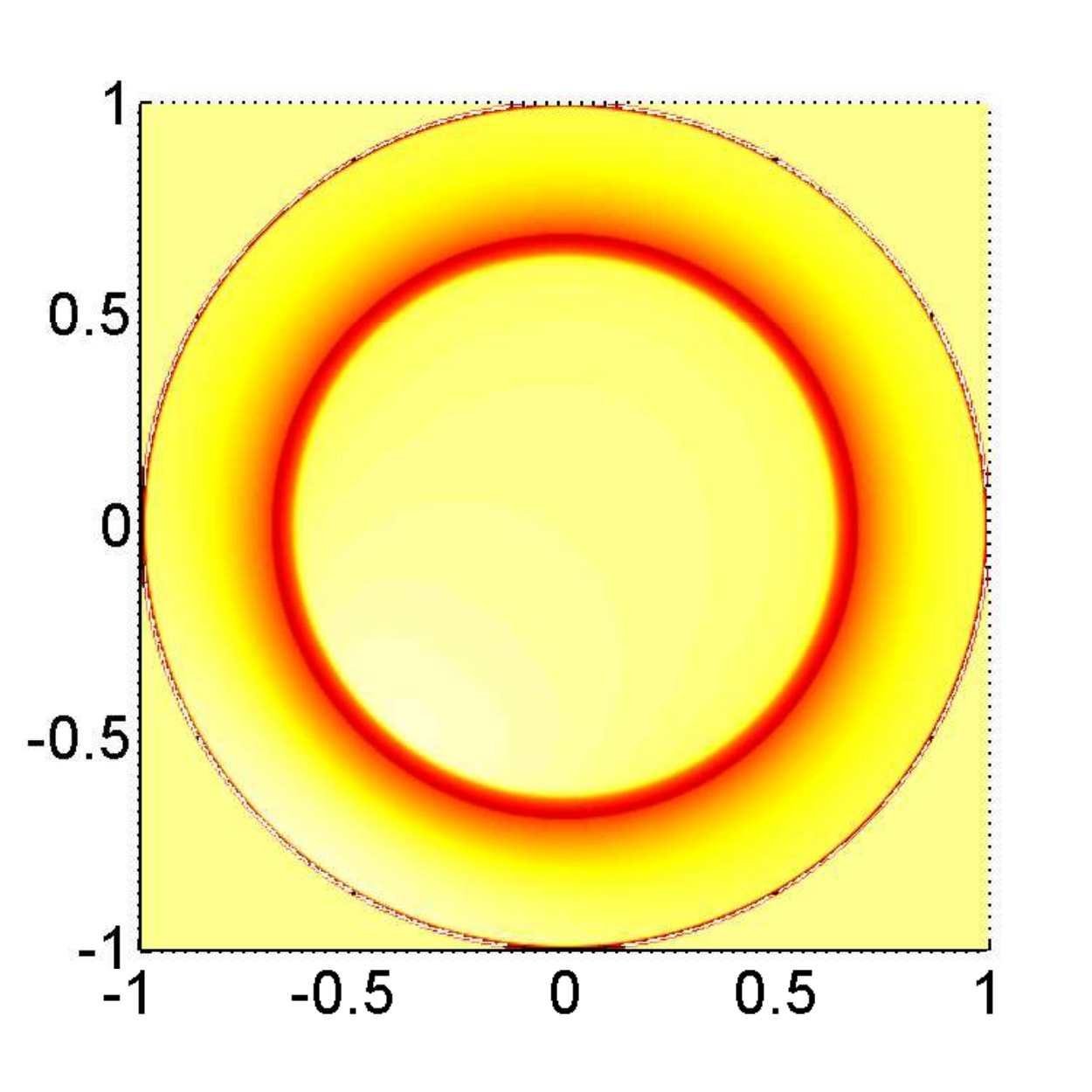}}
   \subfigure[t=10]{\includegraphics[height=1.5in]{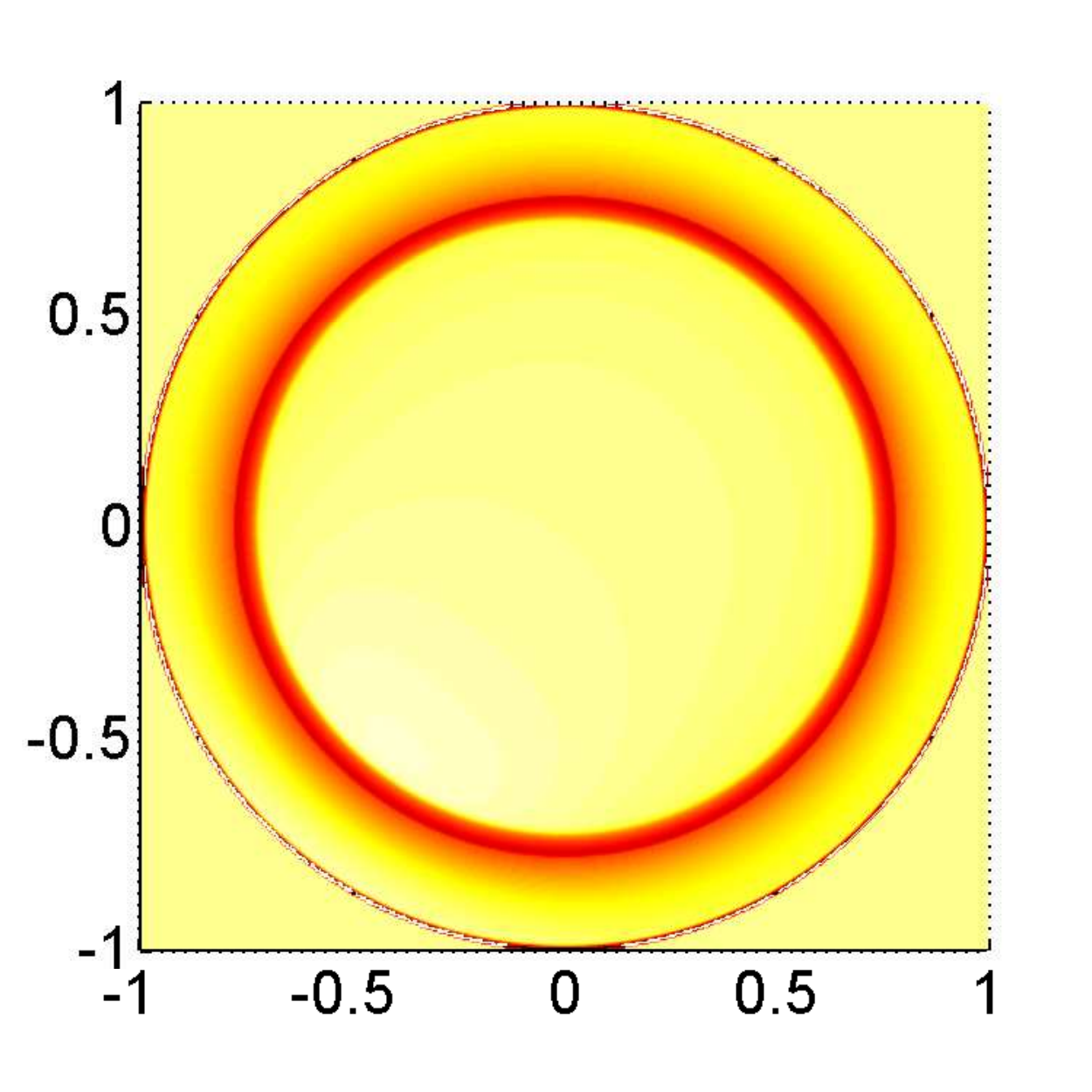}\label{if-i}}
   \caption{{\bf{Simulation of tumor invasion with nonlinear model.}} This figure shows simulations (using (\ref{eq:bpa})) of invasion of a spherically symmetric tumor.  The parameter values are chosen to match qualitatively with the behavior of the invasive rim in figure \ref{stein}.}
   \label{bpa}
 \end{figure}

 Based on the scratch assay experiments and the nonlinear model for cell motility presented above we modify the model of Stein et al.\ (\ref{eq:stein})
 to have the form
 \begin{align}
    \partial_{t}u & = \nabla \cdot (D(u)\nabla u) + s\delta(x - x_{0} - {\bf v} t ) + gu\left(1-\frac{u}{u_{\text{max}}}\right), \label{eq:bpa}
 \end{align}
 where the motility is described by a nonlinear diffusion term of the general form $\nabla \cdot (D(u)\nabla u)$. We interpret this as having the
 motility dependent on the population density. This is of particular interest in the case where there is cell-to-cell adhesion. Figure \ref{bpa}
 shows simulations using the model (\ref{eq:bpa}). We note the qualitative similarity of the behavior of the invasive rim. One difference however, is that we do not treat the cells at the tumor core separately. This is the reason for the difference in centers of the simulations. Figure \ref{bpa}
 does not have the ``hole'' in the center that figure \ref{stein} does.

 In conclusion, based on the data from the scratch assays performed in the Stipp Lab we are able to obtain information on the random walks
 of individual cells. From the individual random walks we are able to classify the results of the experiments in terms of the diffusion and
 drift coefficients. Using these coefficients we can represent the motility of cells by a linear drift-diffusion model. This model is capable
 of providing a qualitative match to experimental observations. Further, by using a nonlinear diffusion model, we can again represent
 the motility of cells such that we again obtain a qualitative match to experimental observations. The value of the nonlinear model is that
 it includes motility that is population density dependent, which is relevant in the case that there is cell-cell adhesion. Moreover, for certain applications such as tumor invasion the nonlinear model may be appropriate since as discussed in \cite{friedl} cell-cell adhesion and hence
 population density dependent migration occurs. The models developed here provide an alternative for capturing directed cell motility in cases
 where mechanisms such as chemotaxis, haptotaxis, etc.\ are not dominant, and when there are no external factors such as anisotropic tissue to influence the direction of cell motion.

 \section*{Acknowledgments} We would like to thank Chris Stipp, Michael Henry, and Justin Drake for numerous discussions on scratch assays and the biology of cell motility. We also thank Lihe Wang and Tong Li for valuable discussions on nonlinear PDEs.

 \vspace{0.3in}
 \begin{flushleft}
 Jason M.\ Graham, Bruce P.\ Ayati \\
 University of Iowa Department of Mathematics \\
 14 MacLean Hall \\
 Iowa City, IA 52242-1419 \\
 {\tt jason-graham@uiowa.edu}\\
 {\tt bruce-ayati@uiowa.edu}
 \end{flushleft}

\newpage

\bibliography{JMGthesis}
\bibliographystyle{plain}

\end{document}